\documentclass[10pt,a4paper,english]{article}
\usepackage{color}
\usepackage[T1]{fontenc}
\usepackage[utf8]{inputenc}
\usepackage{amsfonts}
\usepackage{amssymb}
\usepackage{mathrsfs}
\usepackage{amsthm}
\usepackage{graphicx}
\usepackage{amsmath}
\newcommand{\BH}{{\mathcal{B}(\mathcal{H})}}
\newcommand{\BK}{{\mathcal{B}(\mathcal{K})}}
\newcommand{\BHp}{{\mathcal{B}_+(\mathcal{H})}}
\newcommand{\BKp}{{\mathcal{B}_+(\mathcal{K})}}

\newcommand{\ra}{{\, \rightarrow\, }}
\newtheorem{thm}{Theorem}

\newtheorem{prop}{Proposition}
\newtheorem{cor}{Corollary}
\newtheorem{lemma}{Lemma}
\theoremstyle{definition}

\newcommand{\Tr}[0]{\mathrm{Tr}}
\newcommand{\ot}[0]{\otimes}
\newcommand{\bei}{\begin{itemize}}
\newcommand{\eei}{\end{itemize}}
\newcommand{\ket}[1]{|#1\rangle}
\newcommand{\bra}[1]{\langle#1|}

\newcommand{\proj}[1]{\ket{#1}\!\bra{#1}}
\newcommand{\braket}[2]{\langle{#1}|{#2}\rangle}

\newcommand{\spa}[0]{\mathrm{span}}
\newcommand{\In}{{\mathbb{I}_n}}
\newcommand{\IIn}{{\mathbb{I}_{2n}}}
\newcommand{\On}{{\mathbb{O}_n}}
\def\<{\langle}
\def\>{\rangle}
\newcommand{{\Cn}}{{\mathbb{C}^n}}
\newcommand{{\CN}}{{\mathbb{C}^{2n}}}
\newcommand{{\BC}}{{\mathcal{B}(\mathbb{C}^n)}}
\newcommand{{\BBC}}{{\mathcal{B}(\mathbb{C}^{2n})}}
\def\oper{{\mathchoice{\rm 1\mskip-4mu l}{\rm 1\mskip-4mu l}{\rm 1\mskip-4.5mu l}{\rm 1\mskip-5mu l}}}
\date{}

\begin{document}

\title{\textbf{A class of exposed indecomposable positive maps}}

\author{Gniewomir Sarbicki$^{1,2}$ and Dariusz  Chru\'sci\'nski$^1$\\
$^1$ Institute of Physics, Nicolaus Copernicus University,\\
Grudzi\c{a}dzka 5/7, 87--100 Toru\'n, Poland\\
$^2$ Stockholms Universitet, Fysikum, S-10691 Stockholm, Sweden}

\maketitle

\begin{abstract}
Exposed positive maps in matrix algebras define a dense subset of extremal maps.
We provide a class of indecomposable positive maps in the algebra of $2n \times 2n$ complex matrices with $n \geq 2$. It is shown that these maps are exposed and hence define the strongest tool in entanglement theory to discriminate between separable and entangled states.
\end{abstract}


\section{Introduction}

Entanglement is one of the essential features of quantum physics and
is fundamental  in modern quantum technologies \cite{QIT,HHHH}. One of the central problems in the entanglement theory is the discrimination between separable and entangled states. There are several tools which can be used for this purpose. The most general consists in applying the theory of linear positive maps \cite{HHHH,Guhne}

Recall that a linear map $\Lambda : \BK \ra \BH$ is positive if it maps a cone of positive elements in $\BK$ into  a cone of positive elements in $\BH$, that is, $\Lambda(\BHp) \subset \BKp$ \cite{Paulsen}. Consider a quantum state of a system living in $\mathcal{H} \ot \mathcal{K}$ represented by a density operator $\rho$. It  is separable if and only if $(\oper \ot \Lambda)\rho \geq 0$ for all positive maps $\Lambda : \BK \ra \BH\,$ ($\oper$ denotes an identity map in $\BH$, that is, $\oper(X)=X$).  It is therefore clear that the knowledge of positive maps  $\BK \ra \BH$ allows one to classify states of a composed quantum system living in $\mathcal{H}\ot \mathcal{K}$.
Unfortunately, in spite of the considerable effort, the structure of
positive maps is rather poorly understood \cite{Stormer}--\cite{Eom}. For recent papers about positive maps in entanglement theory see e.g. \cite{Lew}--\cite{Kye-OSID}.

Let $\mathcal{P}^+$ denote a convex cone of positive maps $\BK \ra \BH$. There is a natural question: what is the minimal subset of $\mathcal{P}^+$ which allows to discriminate between all separable and entangled states in $\mathcal{H}\ot \mathcal{K}$? Usually one looks for so called optimal maps (see next Section). It is well known that optimal maps allow to detect all entangled states. Could we further reduce this set? It turns out that the answer to this question is positive. The smallest set of maps needed to detect all entangled states is provided by so called exposed maps. It is, therefore, clear that the knowledge of these maps is highly desired.

In this paper we consider a class of positive maps $\Phi_n : \mathcal{B}(\CN) \ra \mathcal{B}(\CN)$.  These maps were already considered in \cite{Justyna1}. It was shown that they are indecomposable and optimal (indecomposability means that $\Phi_n$ can detect entangled states with positive partial transpose). Here we show that for $n\leq 2$ these maps are not only optimal but even exposed.
In general, the construction of exposed maps is highly involved and we know only few examples of such maps (see e.g. \cite{EX1}--\cite{EX3}). This way our paper extends the knowledge of exposed maps.

The paper is structured as follows: in the next section we explain a concept of positive exposed maps.
Section \ref{CLASS} introduces the specific class of maps considered in this paper. In Section \ref{PROOF} we provide the proof of our main result and finally conclude in the last section.

\section{Exposed maps -- preliminaries}

Recall that a map $\Phi : \BK \ra \BH$ is {\em optimal} if for any completely positive map $\Lambda^{\rm CP} : \BK \ra \BH$ a map $\Phi - \Lambda^{\rm CP}$ is no longer positive. How to check that a given positive map $\Phi$ is optimal?  One has the following

\begin{thm}[\cite{Lew}] \label{TH1}
Let $\Phi : \BK \ra \BH$ and let
\begin{equation*}\label{}
    \mathcal{P}_\Phi = {\rm span} \{ \, |x\> \ot |y\> \in \mathcal{K} \ot \mathcal{H}\ : \ \Phi(|x\>\<x|)|y\> = 0 \, \}\ .
\end{equation*}
If $\mathcal{P}_\Phi = \mathcal{K} \ot \mathcal{H}$, that is,
\begin{equation}\label{WEAK}
    {\rm dim }\, \mathcal{P}_\Phi = d_\mathcal{K}d_\mathcal{H}\ ,
\end{equation}
then $\Phi$ is optimal.
\end{thm}
This property is usually referred as the spanning property \cite{Lew}. Now, among optimal maps one has a subset of extremal maps.
Note that each positive map $\Phi$ in $\mathcal{P}^+$ gives rise to the ray $[\Phi] := \{\lambda \Phi :  \lambda >0\}$.
A positive  map $\Phi : \BK \ra \BH$ is {\em extremal} if for any  positive map $\Lambda : \BK \ra \BH$ which does not belong to the ray $[\Phi]$, a map $\Phi - \Lambda$ is no longer positive. In this case one usually calls $[\Phi]$ an extremal ray. Finally, an extremal ray $[\Phi]$ is {\em exposed} if there exists a supporting hyperplane $H$ such that  $H \cap \mathcal{P}^+ = [\Phi]$. A map $\Phi$ is exposed if and only if it generates an exposed ray $[\Phi]$.

Actually, the theory of  exposed maps may be presented in a much more sophisticated way using elegant geometry of convex cones \cite{convex,Kye}.

One may ask a natural question: why exposed maps are important? Due to the Straszewicz theorem one knows that exposed maps are dense in the set of extremal maps.

Now, a linear map $\Phi : \BK \ra \BH$ is called {\em irreducible} if the following condition holds: if $[\Phi(X),Z] =0$ for all $X \in \BK$, then $Z = \lambda \mathbb{I}_\mathcal{H}$. Actually, one may restrict oneself to self-adjoint elements $\mathcal{B}_{\rm sa}(\mathcal{K})$ only. Indeed, suppose that $\Phi$ is irreducible and $[\Phi(X),Z] =0$ for all $X \in \mathcal{B}_{\rm sa}(\mathcal{K})\,$. Any element $X \in \BK$ may be decomposed as $X = X_1 + i X_2$, with $X_1,X_2 \in \mathcal{B}_{\rm sa}(\mathcal{K})$. One has
$$ [\Phi(X),Z] = [\Phi(X_1),Z] + i[\Phi(X_2),Z] = 0 \ , $$
and irreducibility of $\Phi$ implies therefore $Z = \lambda\mathbb{I}_\mathcal{H}$.
In analogy to Theorem \ref{TH1} one proves \cite{EX3} (see also \cite{Wor2,private})

\begin{thm}  \label{TH-EX}
Let $\Phi :  \mathcal{B}(\mathcal{K}) \longrightarrow \mathcal{B}(\mathcal{H})$ be a positive, unital, irreducible map, and let
\begin{equation*}\label{}
    N_\Phi = {\rm span} \{ \, a \ot |h\> \in \mathcal{B}_+(\mathcal{K}) \ot \mathcal{H} \ : \ \Phi(a)|h\> = 0\, \}\ .
\end{equation*}
If the subspace $N_\Phi \subset \mathcal{B}(\mathcal{K}) \ot \mathcal{H}$ satisfies
\begin{equation}\label{STRONG}
D := {\rm dim}\, N_\Phi =  (d_\mathcal{K}^2 -1)d_\mathcal{H}\ ,
\end{equation}
then $\Phi$ is exposed.
\end{thm}

In analogy to (\ref{WEAK}) we proposed \cite{EX3} to call (\ref{STRONG}) a {\em strong spanning property}.

\section{A class of exposed maps}   \label{CLASS}

In this section we introduce the class of positive maps that we are going to analyze. The starting point is the well known reduction map in $\mathcal{B}(\mathbb{C}^2)$ defined by
\begin{equation}\label{}
    R_2(X) = \mathbb{I}_2 \, {\rm Tr}\, X - X \ ,
\end{equation}
which is known to be a unital, extremal (and hence optimal) positive decomposable map. Actually, it turns out that $R_2$ is even exposed \cite{EX3}. This may be easily generalized
to a linear map $R_n : \mathcal{B}(\mathbb{C}^n) \ra \mathcal{B}(\mathbb{C}^n)$ as follows
\begin{equation}\label{Rn}
    R_n(X)= \mathbb{I}_{n}\mathrm{Tr}X-X \ ,
\end{equation}
for $X \in \mathcal{B}(\mathbb{C}^n)$. Note, that $R_n$ is a positive decomposable map.
However, if $n > 2$ it is no longer extremal and hence can not be exposed. Using $R_n$  let us define the following linear map $\Phi_n:\mathcal{B}(\mathbb{C}^{2n}) \to \mathcal{B}(\mathbb{C}^{2n})$
\begin{equation}\label{R4}
    \Phi_n\left( \begin{array}{c|c} X_{11} & X_{12} \\ \hline X_{21} & X_{22} \end{array} \right) =
\frac 1n \left( \begin{array}{c|c} \mathbb{I}_n \mbox{Tr} X_{22} & -X_{12} - R_n(X_{21}) \\ \hline -X_{21} - R_n(X_{12}) & \mathbb{I}_n\, \mbox{Tr} X_{11} \end{array} \right)\
=: \frac 1n \left( \begin{array}{c|c} Y_{11} & Y_{12} \\ \hline Y_{21} & Y_{22} \end{array} \right)
,
\end{equation}
Note that using $\CN = \Cn \oplus \Cn$,  the blocks $X_{kl}, \, Y_{kl}$ can be perceived as elements of $\BC$. One has $\Phi_n (\mathbb{I}_{2n}) = \mathbb{I}_{2n}$ and it was proved \cite{Justyna1} that $\Phi_n$ defines a positive map.

The main result of this paper consists in the following

\begin{thm}  \label{MAIN}
$\Phi_n$ is exposed for all $n$.
\end{thm}

To prove the above theorem we shall use Theorem \ref{TH-EX}, that is, we prove that $\Phi_n$ satisfies the following two propositions:

\begin{prop}   \label{prop-I}
$\Phi_n$ is irreducible.
\end{prop}

\begin{prop}   \label{prop-II}
The corresponding linear subspace $N_{\Phi_n}$  spanned by vectors $$\overline{x} \ot x \ot y \in \CN \ot \CN \ot \CN\ , $$
such that
$$ \Phi_n(|x\>\<x|)|y\> = 0\ , $$ satisfies
$\ {\rm dim} N_{\Phi_n} = 2n(4n^2-1)$.
\end{prop}

\begin{cor}
Note that $\Phi_2$ reproduces the well known Robertson map \cite{Robertson}.
It was already proved by Robertson that $\Phi_2$ is extremal. Our result shows that being extremal, it is exposed as well.
\end{cor}

\begin{cor}
The following family of maps
\begin{equation}\label{}
 \Phi_n^{U,V}(X) :=   V^* \Phi_n(UXU^*) V\ ,
\end{equation}
is unital and exposed for any unitaries $U,V : \CN \ra \CN\,$.
\end{cor}


\begin{cor} If we relax unitality, then
\begin{equation}\label{}
 \Phi_n^{A,B}(X) :=   A^* \Phi_n(BXB^*) A\ ,
\end{equation}
is  exposed for any $A,B \in GL(2n)\,$.
\end{cor}

\section{Proof of the main result}
\label{PROOF}

\subsection{Proof of Proposition \ref{prop-I}}

We look for $Z \in \BBC$ such that $[\Phi_n(X),Z]=0$ for all $X \in \BBC$. Let us denote $Y:= n\Phi_n(X)$. Condition
\begin{displaymath}
 \left[ \begin{array}{cc} Y_{11} & Y_{12} \\ \nonumber Y_{21} & Y_{22}\end{array} \right] \cdot
 \left[ \begin{array}{cc} Z_{11} & Z_{12} \\ \nonumber Z_{21} & Z_{22}\end{array} \right] =
 \left[ \begin{array}{cc} Z_{11} & Z_{12} \\ \nonumber Z_{21} & Z_{22}\end{array} \right] \cdot
 \left[ \begin{array}{cc} Y_{11} & Y_{12} \\ \nonumber Y_{21} & Y_{22}\end{array} \right]
\end{displaymath}
implies
\begin{align}
 & Y_{12} Z_{21} = Z_{12} Y_{21} \label{kom1}\ , \\
 & Y_{21} Z_{12} = Z_{21} Y_{12} \label{kom2} \ ,\\
 & Y_{21} Z_{11} + Y_{22} Z_{21} = Z_{21} Y_{11} + Z_{22} Y_{21}\ , \label{kom3} \\
 & Y_{11} Z_{12} + Y_{12} Z_{22} = Z_{11} Y_{12} + Z_{12} Y_{22} \ .\label{kom4}
\end{align}
 Let $X$ be block-diagonal, that is, $X_{12}=X_{21}=0$.
Note that $\Phi_n$ maps block-diagonal matrices into block-diagonal matrices and hence
equations (\ref{kom1}), (\ref{kom2}) are trivially  satisfied ($Y_{12}=Y_{21}=0$) and equations (\ref{kom3}) and (\ref{kom4}) imply
\begin{align*}
 & Z_{21}\Tr X_{11}  = Z_{21}\Tr X_{22} \ ,  \\
 & Z_{12}\Tr X_{11}  = Z_{12} \Tr X_{22} \ .
\end{align*}
 Now, due to the fact that $\Tr X_{11}$ and $\Tr X_{22}$ are arbitrary, one has $Z_{12}=Z_{21}=0$, and hence  equations (\ref{kom3}) and (\ref{kom4}) reduce to
\begin{align*}
 & Y_{21} Z_{11} = Z_{22} Y_{21}\ , \\
 & Y_{12} Z_{11} = Z_{22} Y_{12}\ .
\end{align*}
Taking $X_{12}=X_{21}=\In$, one gets $Y_{12} = Y_{21} = -n\In$ and hence $Z_{11}=Z_{22} =: Z_0$. Finally, one obtains the following condition for the diagonal block $Z_0$:
$$ [X_{12} - X_{21},Z_0] = 0\ , $$
%
and since $X_{12}$ and $X_{21}$ are arbitrary, it implies $Z_0 = c\, \In$ and hence $Z = c\, \IIn$, which ends the proof of irreducibility of $\Phi_n$.  \hfill  $\Box$

\subsection{Proof of Proposition \ref{prop-II}}

Before we prove Proposition \ref{prop-II} we need few additional results.
Let $\Psi = \sum_{i,j,k} \Psi_{ijk}\, e_i \ot e_j \ot e_k \in \mathbb{C}^{n} \ot \mathbb{C}^{n} \ot \mathbb{C}^{n}$.
We define the following subspaces in $(\mathbb{C}^n)^{\otimes 3}$: let $S_{123}$ be a totally symmetric subspace,
i.e. $\Psi \in S_{123}$ iff $\Psi_{ijk} = \Psi_{\pi(i)\pi(j)\pi(k)}$ for an arbitrary permutation $\pi$.
Moreover, let us introduce
\begin{eqnarray*}
 \Psi \in S_{23}  & \ \mbox{if}\  &\ \Psi_{ijk} = \Psi_{ikj} \ , \\
 \Psi \in A_{23}  & \ \mbox{if}\ & \Psi_{ijk} = -\Psi_{ikj} \ , \\
 \Psi \in T_{13}  & \ \mbox{if}\ & \sum_{i} \Psi_{iji} = 0 \ , \\
 \Psi \in I_{13}  & \ \mbox{if}\ & \Psi_{ijk} = \lambda_{j} \delta_{ik} \ .
\end{eqnarray*}
One easily finds for the corresponding dimensions
\begin{eqnarray} \label{dimSATI}
 \dim S_{23} = \frac{n^2(n+1)}2 \ , \ \ \
 \dim A_{23} = \frac{n^2(n-1)}2 \ ,
\end{eqnarray}
and
\begin{eqnarray} \label{dimSATII}
 \dim T_{13} = n(n^2-1) \ , \ \ \
 \dim I_{13} = n  \ .
\end{eqnarray}

\begin{lemma} \label{ST}
 Any element of $(\mathbb{C}^n)^{\otimes 3}$ can be decomposed as a sum of an element from $S_{123}$ and an element from $T_{13}\,$.
\end{lemma}

\proof Let $\Psi$ be an arbitrary element of $(\mathbb{C}^n)^{\otimes 3}$. We define $A,B\in (\mathbb{C}^n)^{\otimes 3}$
$$ A = \sum_{i,j,k} A_{ijk}\, e_i \ot e_j \ot e_k\ , \ \ \ B = \sum_{i,j,k} B_{ijk}\, e_i \ot e_j \ot e_k\ , $$
as follows:
\begin{align*}
 & A_{ijk} = \left\{ \begin{array}{ll} \sum_m \Psi_{mim} & \mathrm{if\ } i=j=k \\
              0 & \mathrm{otherwise}
             \end{array}
	     \right.
\end{align*}
and
\begin{align*}
 & B_{ijk} = \left\{ \begin{array}{ll} \Psi_{ijk} - \sum_m \Psi_{mim} & \mathrm{if\ } i=j=k \\
              \Psi_{ijk} & \mathrm{otherwise}
             \end{array} \right.
\end{align*}
Clearly $A+B=\Psi$, $A \in S_{123}$ and $B \in T_{13}\,$. \hfill  $\Box$

We stress, that this decomposition is not unique.
Actually, since $S_{123} \subset S_{23}$ one has $A \in S_{23}$ and hence
\begin{equation}\label{S23-T13}
    (\mathbb{C}^n)^{\otimes 3} = (S_{23} + T_{13})\ .
\end{equation}
In what follows we use the following notation: by $W + V$ we denote the a set of vectors $w + v$, where $w\in W$ and $v\in V$. Note, that it differs from the direct sum $W \oplus V$, where the decomposition $w + v$ is unique.

\vspace{0.3cm}

\paragraph{Proof of Proposition \ref{prop-II}}

Now, we are ready to prove Proposition \ref{prop-II}.
Let $P=|x\>\<x|$ with arbitrary $x \in \mathbb{C}^{2n}$. Now, due to
$\CN=\Cn \oplus \Cn$, one has $x = \varphi_1 \oplus \varphi_2$ and hence $P$ displays the following block structure
$$ P = \left(   \begin{array}{c|c}
  X_{11} & X_{12} \\ \hline  X_{21} & X_{22} \end{array}   \right)
\ , $$
with $X_{ij} = |\varphi_i\>\<\varphi_j|$. We get
\begin{equation}\label{}
    \Phi_n(P) =
    \frac 1n \left( \begin{array}{c|c} ||\varphi_2||^2 \In &  -\ket{\varphi_1}\bra{\varphi_2} + \ket{\varphi_2}\bra{\varphi_1}-\braket{\varphi_1}{\varphi_2}\, \In \\ \hline
      -\ket{\varphi_2}\bra{\varphi_1} + \ket{\varphi_1}\bra{\varphi_2}-\braket{\varphi_2}{\varphi_1}\, \In & ||\varphi_1||^2 \In \end{array}
\right)\ .
\end{equation}
We are looking for vectors $y = \psi_1 \oplus \psi_2$, such that $\bra{y}\Phi_n(\proj{x})\ket{y}=0$. Observe, that
 \begin{align}
  & n\Phi_n(\proj{x}) = \left( \begin{array}{cc} ||\varphi_2||^2 & -\braket{\varphi_1}{\varphi_2} \\ -\braket{\varphi_2}{\varphi_1} & ||\varphi_1||^2 \end{array} \right) \otimes \In \\ \nonumber
  & + \left( \begin{array}{c|c} \On & -\ket{\varphi_1}\bra{\varphi_2} + \ket{\varphi_2}\bra{\varphi_1} \\ \hline -\ket{\varphi_2}\bra{\varphi_1} + \ket{\varphi_1}\bra{\varphi_2} & \On \end{array} \right) \ , \nonumber
 \end{align}
where $\mathbb{O}_n$ denotes the $n\times n$ matrix with all elements equal to zero.
Now, the first term is strictly positive iff the vectors $\varphi_1$ and  $\varphi_2$ are not parallel.  The second term acts only in subspace $\spa\{\varphi_1,\varphi_2\}$. It is therefore clear that to have $\bra{y}\Phi(\proj{x})\ket{y} = 0$  for some $y=\psi_1\oplus\psi_2$ at least one of the following conditions has to be satisfied:
\begin{enumerate}
 \item $\varphi_1$ and $ \varphi_2$ are parallel,
 \item $\psi_1, \psi_2 \in \spa\{\varphi_1,\varphi_2\}\,$.
\end{enumerate}

\noindent
We shall consider two cases:
i) $\varphi_1 \parallel \varphi_2\,$, and
ii)  $\varphi_1 \perp \varphi_2\,$. Of course in general $\varphi_1$ and $\varphi_2$ are neither parallel nor perpendicular. However, it turns out that it is sufficient to analyze only these two cases.  Let us introduce the following notation:
\begin{equation}\label{}
     \spa \{\, x^* \otimes x \otimes y: \ \Phi_n(|x\>\<x|)|y\> = 0\, \} \ =:\  \left\{ \begin{array}{ll} W \ , &  {\rm if}\  \varphi_1 \parallel \varphi_2  \\
     V \ , &  {\rm if}\  \varphi_1 \perp \varphi_2 \end{array} \right. \ .
\end{equation}
We shall characterize $W$ and $V$ by providing the basis for $W^\perp$ and $V^\perp$. It is clear that to prove the proposition it is sufficient to show that
\begin{equation}\label{}
    {\rm dim}\, (W + V) \geq 8n^3-2n\ .
\end{equation}

\paragraph{The subspace $W$}

Let us assume $\varphi_1 \parallel \varphi_2$, that is, $\varphi_1=\alpha\, \varphi, \varphi_2 = \beta\, \varphi$ where $\alpha, \beta \in \mathbb{C}$ and  we assume that $\varphi \in \Cn$ satisfies $||\varphi||=1$. For later convenience let us denote $\sigma=[\alpha,\beta]^t, x=\sigma\otimes\varphi$. One obtains
\begin{align} \label{nPhi}
& n\Phi_n(\proj{x}) = \left[ \begin{array}{cc} |\beta|^2 & -\beta \alpha^* \\ -\beta^* \alpha & |\alpha|^2 \end{array} \right] \otimes \In \\ \nonumber  & +
 \left[ \begin{array}{cc} 0 & -\beta^* \alpha+\beta \alpha^* \\ -\beta \alpha^*+\beta^* \alpha & 0 \end{array} \right] \otimes \proj{\varphi}
 \\ \nonumber  & =
 \Pi_{[\beta, -\alpha]^t} \In + (\Pi_{[ \beta^*, -\alpha^*]^t} - \Pi_{[ \beta, -\alpha]^t}) \Pi_\varphi \\ \nonumber  & =
 \Pi_{[\beta, -\alpha]^t} \otimes \Pi_\varphi^\perp + \Pi_{[\beta^*, -\alpha^*]^t} \otimes \Pi_\varphi
 \end{align}
We are looking for elements of the kernel of this operator. Let us change the basis $\{ e_\alpha\}$ in the second factor of the tensor product such that $\phi = e_1$. Then an arbitrary vector $y$ can be decomposed as $\gamma_1 \otimes \phi + \sum_{\alpha=2}^n \gamma_\alpha \otimes e_\alpha$ and the matrix (\ref{nPhi}) is block-diagonal with 2-dimensional blocks $\Pi_{[\beta^*, -\alpha^*]^t}, \Pi_{[\beta, -\alpha]^t}, \dots, \Pi_{[\beta, -\alpha]^t}$. The vector $y$ is in kernel iff the vectors $\gamma_\alpha$ are in kernels of corresponding blocks.
The kernel of $\Phi_n(\proj{\sigma\otimes\varphi})$ is therefore spanned by vectors of the following form
\begin{align}
 & y = \sigma \otimes \varphi \\
 & y = \overline{\sigma} \otimes \varphi^\perp\ .
\end{align}
We are looking for the dimension of subspace $W$ in $(\mathbb{C}^{2n})^{\otimes 3}$ spanned by the vectors
$\, \overline{x} \ot x \ot y\,$ such that $\,\Phi_n(|x\>\<x|)|y\>=0\,$.
 Then $W = W_1 + W_2\,$, where
$$ W_1 = {\rm span}\, \{ \overline{\sigma} \ot \overline{\varphi} \ot \sigma \ot \varphi \ot \sigma \ot \varphi \} \ , $$
and
$$  W_2 = {\rm span}\, \{ \overline{\sigma} \ot \overline{\varphi} \ot \sigma \ot \varphi \ot \overline{\sigma} \ot \varphi^\perp \}\ ,$$
where $\varphi^\perp$ is an arbitrary vector in $\Cn$ orthogonal to $\varphi$.
\begin{lemma} \label{dimW}
  The following statements holds:
 \begin{enumerate}
  \item $\dim W_1=3n^2(n+1)$
  \item $\dim W_2=6n(n^2-1)$
  \item $\dim W = 7n^3+n^2-2n$
  \item the subspace $W^\perp$ is spanned by $n^2(n-1)$ elements:
  \begin{align}\label{u-ijk}
    u_{ijk} & =
   \ket{e_i \oplus 0|e_j \oplus 0|0 \oplus e_k} - \ket{e_i \oplus 0|e_k \oplus 0|0 \oplus e_j} \\
  &  - \ket{0 \oplus e_i|e_j \oplus 0|e_k \oplus 0} + \ket{0 \oplus e_i|e_k \oplus 0|e_j \oplus 0} \ ,
    \end{align}
   and
 \begin{align}\label{v-ijk}
    v_{ijk} & =
   \ket{0 \oplus e_i|0 \oplus e_j|e_k \oplus 0} - \ket{0 \oplus e_i|0 \oplus e_k|e_j \oplus 0} \\
   & - \ket{e_i \oplus 0|0 \oplus e_j|0 \oplus e_k} + \ket{e_i \oplus 0|0 \oplus e_k|0 \oplus e_j}\ ,
  \end{align}
  where $1 \le j<k \le n$ and $1 \le i \le n\,$,
  and additional $2n$ elements:
  \begin{align*}
  & r_i = \sum_{j=1}^n (\ket{e_j \oplus 0|0 \oplus e_i|e_j \oplus 0} - \ket{e_j \oplus 0|e_i \oplus 0|0 \oplus e_j}) \\
  & s_i = \sum_{j=1}^n (\ket{0 \oplus e_j|e_i \oplus 0|0 \oplus e_j} - \ket{0 \oplus e_j|0 \oplus e_i|e_j \oplus 0})
  \end{align*}
  where $1 \le i \le n$.
 \end{enumerate}
\end{lemma}
\proof
It will be convenient to rearrange the factors of the tensor product as $(\mathbb{C}^{2n})^{\ot 3} = (\mathbb{C}^2)^{\otimes 3} \otimes (\mathbb{C}^n)^{\otimes 3}$. Now we introduce the basis $\{f_k\}$ and the dual basis $\{ f_l^* \}$ in $(\mathbb{C}^2)^{\otimes 3}$:
\begin{align} \label{baza}
 & f_1 = \ket{000} \ , & f^*_1 = \ket{000} \ , \nonumber\\
 & f_2 = \ket{111} \ , & f^*_2 = \ket{111} \ ,\nonumber\\
 & f_3 = \frac 1{\sqrt{3}}(\ket{001}+\ket{100}+\ket{010}) \ ,  & f^*_3 = \frac 1{\sqrt{3}}(\ket{001}+\ket{100}+\ket{010}) \ , \nonumber\\
 & f_4 = \frac 1{\sqrt{3}}(\ket{110}+\ket{011}+\ket{101}) \ , & f^*_4 = \frac 1{\sqrt{3}}(\ket{110}+\ket{011}+\ket{101}) \ , \nonumber\\
 & f_5 = \frac 1{\sqrt{6}}(\ket{010}+\ket{001}-2\ket{100}) \ , & f^*_5 = \frac 1{\sqrt{2}}(\ket{001}-\ket{100}) \ , \nonumber \\
 & f_6 = \frac 1{\sqrt{6}}(\ket{101}+\ket{110}-2\ket{011}) \ , & f^*_6 = \frac 1{\sqrt{2}}(\ket{110}-\ket{011})\ ,\nonumber\\
 & f_7 = \frac 1{\sqrt{6}}(\ket{001}+\ket{100}-2\ket{010}) \ ,  & f^*_7 = \frac 1{\sqrt{2}}(\ket{010}-\ket{001}) \ , \nonumber\\
 & f_8 = \frac 1{\sqrt{6}}(\ket{110}+\ket{011}-2\ket{101}) \ , & f^*_8 = \frac 1{\sqrt{2}}(\ket{101}-\ket{110})\ .  \nonumber
\end{align}
As usual the duality of $\{f_k\}$ and $\{ f_l^* \}$  is defined by the following relation
$\< f_k^* | f_l\> = 0$ for $k \neq l$.
Let us introduce the following subspaces in $\mathbb{C}^2 \ot \mathbb{C}^2 \ot \mathbb{C}^2$:
\begin{itemize}
 \item $V_0 = V_0^* = \spa \{f_1, \dots, f_4\}$
 \item $V_1=\spa \{f_5, f_6\}$, $V_1^*=\spa \{f_5^*, f_6^*\}$
 \item $V_2=\spa \{f_7, f_8\}$, $V_2^*=\spa \{f_7^*, f_8^*\}$
\end{itemize}
Observe, that  $V_0$ is totally symmetric under the permutations of all three factors in $\mathbb{C}^2 \ot \mathbb{C}^2 \ot \mathbb{C}^2$. The subspace $V_0 \oplus V_1$ is  invariant under the permutation of  2nd and 3rd  factors and the subspace $V_0 \oplus V_2$ is  invariant under the permutation of the 1st and 3rd  factors in $\mathbb{C}^2 \ot \mathbb{C}^2 \ot \mathbb{C}^2$.

The subspace $W_1$ is generated by elements $(\overline{\sigma} \otimes \sigma \otimes \sigma) \otimes (\overline{\varphi} \otimes \varphi \otimes \varphi)$ and hence
\begin{equation} \label{W1struct}
 W_1 = (V_0 \oplus V_1) \otimes S_{23} \ .
\end{equation}
The subspace $W_2$ is generated by elements $(\overline{\sigma} \otimes \sigma \otimes \overline{\sigma}) \otimes (\overline{\varphi} \otimes \varphi \otimes \varphi^\perp)$. It is again a tensor product of two subspaces. The elements $(\overline{\sigma} \otimes \sigma \otimes \overline{\sigma})$ generate the subspace  $V_0 \oplus V_2$.
The subspace in $(\mathbb{C}^n)^{\otimes 3}$ generated by elements $\overline{\varphi} \otimes \varphi \otimes \varphi^\perp$ reproduces $T_{13}$. Indeed, the matrix corresponding to $\overline{\varphi} \otimes \varphi^\perp$ is traceless. Hence
\begin{equation} \label{W2struct}
 W_2 = (V_0 \oplus V_2) \otimes T_{13}\ .
\end{equation}
Using equations (\ref{dimSATI}) and (\ref{dimSATII}) one gets $\dim W_1 = 3n^2(n+1)$ and $\dim W_2 = 6n(n^2-1)$.

Now, using the decompositions (\ref{W1struct}), (\ref{W2struct}) and the linear dependence of the subspaces $V_0, V_1, V_2$ one finds that $W = W_1 + W_2$ is equal to
\begin{align}
& V_1 \otimes S_{23} \oplus V_0 \otimes (S_{23} + T_{13}) \oplus V_2 \otimes T_{13} = \nonumber \\ & V_1 \otimes S_{23} \oplus V_0 \otimes  (\mathbb{C}^n)^{\otimes 3} \oplus V_2 \otimes T_{13}\ ,
\end{align}
where we have used (\ref{S23-T13}). The dimension of $W$ is then equal to $$2 \cdot \frac{n^2(n+1)}2 + 4 \cdot n^3 + 2 \cdot n(n^2-1)=7n^3+n^2-2n\ , $$ which proves the third point of the lemma.

Let us observe, that any vector orthogonal to $V_1 \otimes S_{23} \oplus V_0 \otimes (\mathbb{C}^n)^{\otimes 3} \oplus V_2 \otimes T_{13}$ has to belong to the subspace $V_0^\perp \otimes (\mathbb{C}^n)^{\otimes 3}$. Now, since $V_0^\perp$ can be decomposed as $V_1^* \oplus V_2^*$, any vector in $V_0^\perp \otimes (\mathbb{C}^n)^{\otimes 3}$ can be decomposed into two parts: $x \in V_1^* \otimes (\mathbb{C}^n)^{\otimes 3}$ and $y \in V_2^* \otimes (\mathbb{C}^n)^{\otimes 3}$. The vector $x \in W^\perp$ iff $x \in V_1^* \otimes A_{23}$. The vector $y \in W^\perp$ iff $y \in V_2^* \otimes I_{23}$.

Let $\{e_i \otimes e_j \otimes e_k - e_i \otimes e_k \otimes e_j: \ j<k \}$ be the basis of $A_{23}$.  The basis of $V_1^* \otimes A_{23}$ contains two families of vectors:
\begin{align*}
 & u_{ijk} = f_5^* \otimes (e_i \otimes e_j \otimes e_k - e_i \otimes e_k \otimes e_j) \\ & = (\ket{001}-\ket{100}) \otimes (e_i \otimes e_j \otimes e_k - e_i \otimes e_k \otimes e_j)\ ,
\end{align*}
and
\begin{align*}
 & v_{ijk} = f_6^* \otimes (e_i \otimes e_j \otimes e_k - e_i \otimes e_k \otimes e_j)\\ & = (\ket{110}-\ket{011}) \otimes (e_i \otimes e_j \otimes e_k - e_i \otimes e_k \otimes e_j)
\end{align*}
Rearranging  the factors of the tensor product, one finds:
\begin{align*}
  u_{ijk} & = (0 \otimes e_i) \otimes (0 \otimes e_j) \otimes (1 \otimes e_k)  \\ & -
 (0 \otimes e_i) \otimes (0 \otimes e_k) \otimes (1 \otimes e_j) \\ & -
 (1 \otimes e_i) \otimes (0 \otimes e_j) \otimes (0 \otimes e_k) \\ & +
 (1 \otimes e_i) \otimes (0 \otimes e_k) \otimes (0 \otimes e_j) \\
 &  =:  \ket{e_i \oplus 0|e_j \oplus 0|0 \oplus e_k} - \ket{e_i \oplus 0|e_k \oplus 0|0 \oplus e_j} \\ & -
 \ket{0 \oplus e_i|e_j \oplus 0|e_k \oplus 0} + \ket{0 \oplus e_i|e_k \oplus 0|e_j \oplus 0} \ ,
\end{align*}
and similarly
\begin{align*}
  v_{ijk} & =
   \ket{0 \oplus e_i|0 \oplus e_j|e_k \oplus 0} - \ket{0 \oplus e_i|0 \oplus e_k|e_j \oplus 0}\\ & - \ket{e_i \oplus 0|0 \oplus e_j|0 \oplus e_k} + \ket{e_i \oplus 0|0 \oplus e_k|0 \oplus e_j}\ .
\end{align*}
Similarly, let $\{\sum_{j} e_j \otimes e_i \otimes e_j\}$ be the basis of $I_{23}$. The basis of $V_2 \otimes I_{13}$ contains two families of vectors:
\begin{align*}
  r_i &= f_7^* \otimes \sum_{j} e_j \otimes e_i \otimes e_j = (\ket{010}-\ket{001}) \otimes \sum_{j} e_j \otimes e_i \otimes e_j \ ,\\
  s_i & = f_8^* \otimes \sum_{j} e_j \otimes e_i \otimes e_j = (\ket{101}-\ket{110}) \otimes \sum_{j} e_j \otimes e_i \otimes e_j\ .
\end{align*}
Rearranging  the factors of the tensor product, one gets:
\begin{align*}
  r_i & = \sum_j[ (0\otimes e_j)\otimes(1 \otimes e_i)\otimes(0 \otimes e_j) -(0\otimes e_j)\otimes(0 \otimes e_i)\otimes(1 \otimes e_j)]  \\ &= \sum_j(\ket{e_j \oplus 0|0 \oplus e_i|e_j \oplus 0}-\ket{e_j \oplus 0|e_i \oplus 0|0 \oplus e_j}) \ ,\\
  s_i & = \sum_j[(1\otimes e_j)\otimes(0 \otimes e_i)\otimes(1 \otimes e_j) -(1\otimes e_j)\otimes(1 \otimes e_i)\otimes(0 \otimes e_j)] \\ & = \sum_j(\ket{0 \oplus e_j|e_i \oplus 0|0 \oplus e_j}-\ket{0 \oplus e_j|0 \oplus e_i|e_j \oplus 0})\ ,
\end{align*}
which ends the proof of the lemma.
\hfill $\Box$

\vspace{.3cm}

\paragraph{The subspace $V$}

Now, let us assume $\varphi_1 \perp \varphi_2$. To simplify notation let $x := \varphi_1$ and $y:= \varphi_2\,$, with $\<x|y\>=0$. One easily finds
 \begin{equation} \label{m1}
 n\Phi_n(|\varphi\>\<\varphi|) =  \left[ \begin{array}{c|c} ||y||^2 \In & -\ket{x}\bra{y}+\ket{y}\bra{x} \\ \hline -\ket{y}\bra{x}+\ket{x}\bra{y} & ||x||^2 \In \end{array} \right]\ ,
 \end{equation}
with $\varphi = x \oplus y$. Now, we look for $z \oplus z' \in \mathbb{C}^{2n}$ such that $\Phi_n(|\varphi\>\<\varphi|)|z\oplus z'\> =0$.
It is clear that a necessary condition for $z \oplus z'$ to belong to the kernel of $\Phi_n(|\varphi\>\<\varphi|)$ is that  $z,z' \in \spa \{ x, y\}$. One has therefore
$$z = z_1 x + z_2 y \ , $$
with $z_1,z_2 \in \mathbb{C}$. We calculate $z'$ using the lower row of blocks in (\ref{m1}). The formula for $z'$ reads
\begin{displaymath}
  z' = -\frac 1{||x||^2} (-\ket{x}\bra{y}+\ket{y}\bra{x}) |z\> = \frac 1{||x||^2} (z_1 ||x||^2 y - z_2 ||y||^2 x)\ .
\end{displaymath}
It finally leads to the following formula (up to an irrelevant complex factor)
$$ z \oplus z' =: [z_1 ||x||^2 x + z_2 ||x||^2 y, z_1 ||x||^2 y - z_2 ||y||^2 x]\ . $$
Now, we look for a subspace $V$ in $\mathbb{C}^{2n} \ot \mathbb{C}^{2n}  \ot \mathbb{C}^{2n} $ spanned by vectors
of the following form:
\begin{equation}\label{zzz}
\Psi =  [x,y]^* \otimes [x,y] \otimes [z_1 ||x||^2 x + z_2 ||x||^2 y, z_1 ||x||^2 y - z_2 ||y||^2 x] \ .
\end{equation}
Let $\{ e^{(1)}_k\}$ and $\{ e^{(2)}_k\}$ denote two orthonormal bases in $\mathbb{C}^n$. Then $e^{(1)}_k \oplus e^{(2)}_l $ defines an orthonormal basis in $\mathbb{C}^{2n} = \mathbb{C}^n \oplus \mathbb{C}^n$ and hence  any vector $\Psi \in \mathbb{C}^{2n} \ot \mathbb{C}^{2n} \ot \mathbb{C}^{2n}$ may be represented as follows:
\begin{equation}\label{}
    \Psi = \sum_{i,j,k=1}^n \sum_{\alpha,\beta,\gamma=1}^2 \Psi^{(\alpha\beta\gamma)}_{ijk} \, e^{(\alpha)}_i  \ot e^{(\beta)}_j  \ot e^{(\gamma)}_k\ .
\end{equation}
One easily finds that coordinates of $\Psi$ defined in (\ref{zzz}) are given by the following polynomial functions:
\begin{align}
\Psi^{(111)}_{ijk} = z_1 ||x||^2 x_i^* x_j x_k + z_2 ||x||^2 x_i^* x_j y_k \ , \nonumber \\
\Psi^{(121)}_{ijk} = z_1 ||x||^2 x_i^* y_j x_k + z_2 ||x||^2 x_i^* y_j y_k \ , \nonumber \\
\Psi^{(211)}_{ijk} = z_1 ||x||^2 y_i^* x_j x_k + z_2 ||x||^2 y_i^* x_j y_k  \ , \nonumber \\
\Psi^{(221)}_{ijk} = z_1 ||x||^2 y_i^* y_j x_k + z_2 ||x||^2 y_i^* y_j y_k  \ , \label{wekt_V} \\
\Psi^{(112)}_{ijk} = z_1 ||x||^2 x_i^* x_j y_k - z_2 ||y||^2 x_i^* x_j x_k \ , \nonumber \\
\Psi^{(122)}_{ijk} = z_1 ||x||^2 x_i^* y_j y_k - z_2 ||y||^2 x_i^* y_j x_k \ , \nonumber \\
\Psi^{(212)}_{ijk} = z_1 ||x||^2 y_i^* x_j y_k - z_2 ||y||^2 y_i^* x_j x_k \ , \nonumber \\
\Psi^{(222)}_{ijk} = z_1 ||x||^2 y_i^* y_j y_k - z_2 ||y||^2 y_i^* y_j x_k\ \nonumber .
\end{align}
To compute the dimension of $V$ one has to check how many of these polynomials are linearly independent. Let us analyze linear combinations of the above 8 families of polynomials.
Observe, that any polynomial being a combination of functions from one row in family (\ref{wekt_V}) is of the form $z_1 f(x,y) + z_2 g(x,y)$, where all monomials in $f(x,y)$ have the same signatures, and all monomials in $g(x,y)$ as well. The signatures of monomials
\footnote{The signature of a monomial is a tuple of exponents of variables, for example $x_i^\alpha x_j^{*\beta} y_k^\mu y_l^{*\nu}$ has the signature $(\alpha,\beta,\mu,\nu)$}
 of $f$'s ang $g$'s for functions from (\ref{wekt_V}) are listed in the table below :
\begin{displaymath}
\begin{array}{r|llllcllll}
 & x & x^* & y & y^* & \quad & x & x^* & y & y^* \\
 \hline
 \Psi^{(111)}_{ijk} & 2 & 3 & 0 & 0 & & 2 & 2 & 1 & 0 \\
 \Psi^{(121)}_{ijk} & 2 & 2 & 1 & 0 & & 1 & 2 & 2 & 0 \\
 \Psi^{(211)}_{ijk} & 3 & 1 & 0 & 1 & & 2 & 1 & 1 & 1 \\
 \Psi^{(221)}_{ijk} & 2 & 1 & 1 & 1 & & 1 & 1 & 2 & 1 \\
 \Psi^{(112)}_{ijk} & 2 & 2 & 1 & 0 & & 2 & 1 & 1 & 1 \\
 \Psi^{(122)}_{ijk} & 1 & 2 & 2 & 0 & & 1 & 1 & 2 & 1 \\
 \Psi^{(212)}_{ijk} & 2 & 1 & 1 & 1 & & 2 & 0 & 1 & 2 \\
 \Psi^{(222)}_{ijk} & 1 & 1 & 2 & 1 & & 1 & 0 & 2 & 2 \\
\end{array}
\end{displaymath}
We can see, that there are two pairs of rows ($(2,5)$ and $(4,7)$) which have the same signatures of their $f$'s, but then $g$'s of rows $(2,7)$ are different. Similarly, two pairs of rows ($(3,5)$ and $(4,6)$) have the same signatures of their $g$'s, but $f$'s of rows $(3,6)$ are different. We observe, that to get a vanishing combination, one has to take combinations of functions from a row with vanishing $f$ or $g$ parts.

To get a vanishing part $f$ or $g$ of a row one has to have linear dependencies among its monomials. In this family it is possible iff  the set of monomials is symmetric under a permutation of indices. Observe, that in the rows $(1,3,6,8)$ the monomials in $f$ have this feature, and in the rows $(2,4,5,7)$ this applies to the monomials in $g$. To kill the part $f$ or $g$ of a row one has to take monomials related by permutation of indices with opposite coeficients, so to consider combinations of functions of the form:
\begin{equation}
    \chi^{(\alpha\beta\gamma)}_{ijk} := \Psi^{(\alpha\beta\gamma)}_{ijk} - \Psi^{(\alpha\beta\gamma)}_{ikj} \ ,
\end{equation}
for $j \neq k$. One finds
\begin{align*}
\chi^{(111)}_{ijk} = z_2 ||x||^2 x_i^* (x_j y_k - x_k y_j) \ , \\
\chi^{(121)}_{ijk} = z_1 ||x||^2 x_i^* (y_j x_k - y_k x_j) \ ,\\
\chi^{(211)}_{ijk} = z_2 ||x||^2 y_i^* (x_j y_k - x_k y_j) \ , \\
\chi^{(221)}_{ijk} = z_1 ||x||^2 y_i^* (y_j x_k - y_k x_j) \ , \\
\chi^{(112)}_{ijk} = z_1 ||x||^2 x_i^* (x_j y_k - x_k y_j) \ , \\
\chi^{(122)}_{ijk} = z_2 ||y||^2 x_i^* (y_j x_k - y_k x_j) \ , \\
\chi^{(212)}_{ijk} = z_1 ||x||^2 y_i^* (x_j y_k - x_k y_j) \ ,\\
\chi^{(222)}_{ijk} = z_2 ||y||^2 y_i^* (y_j x_k - y_k x_j)\ .
\end{align*}
Monomials $\chi^{(\alpha\beta\gamma)}_{ijk}$ containing $z_2$ are linearly independent. However, one has
\begin{equation}\label{}
    \chi^{(121)}_{ijk} = - \chi^{(112)}_{ijk} \ , \ \ \ \chi^{(221)}_{ijk} = - \chi^{(212)}_{ijk} \ .
\end{equation}
It gives therefore $2 \cdot n \cdot \frac{n(n-1)}2$  relations between $\chi$s. They correspond to scalar products of a vector from $V$ with the following vectors
\begin{align*}
 & a_{ijk} = e_i^{(1)} \otimes \Big[ e_j^{(1)} \otimes e_k^{(2)} - e_k^{(1)} \otimes e_j^{(2)} + e_j^{(2)} \otimes e_k^{(1)} - e_k^{(2)} \otimes e_j^{(1)}  \Big] \ , \\
 & b_{ijk} = e_i^{(2)} \otimes \Big[ e_j^{(1)} \otimes e_k^{(2)} - e_k^{(1)} \otimes e_j^{(2)} + e_j^{(2)} \otimes e_k^{(1)} - e_k^{(2)} \otimes e_j^{(1)}  \Big] \ .
\end{align*}
There is no other way to get a vanishing linear combination. Recall, however, that $x \perp y$ and hence any polynomial containing
$\sum_i x_i^* y_i$ or $\sum_i y_i^* x_i$ will vanish as well. Let us compute $\sum_{ij} \delta_{ij} \Psi^{(\alpha\beta\gamma)}_{ijk}$ and
$\sum_{ik} \delta_{ik} \Psi^{(\alpha\beta\gamma)}_{ijk}$. One finds
\begin{eqnarray}\label{ij1}
\sum_{i,j} \delta_{ij} \Psi^{(111)}_{ijk} &=& z_1 ||x||^4 x_k + z_2 ||x||^4 y_k \ , \\ \label{ij2}		
\sum_{i,j} \delta_{ij} \Psi^{(121)}_{ijk} &=& 0 \ ,  \\ \label{ij3}
\sum_{i,j} \delta_{ij} \Psi^{(211)}_{ijk} &=& 0 \ , \\ \label{ij4} 						
\sum_{i,j} \delta_{ij} \Psi^{(221)}_{ijk} &=& z_1 ||x||^2 ||y||^2 x_k + z_2 ||x||^2 ||y||^2 y_k 	\ , \\ \label{ij5}
\sum_{i,j} \delta_{ij} \Psi^{(112)}_{ijk} &=& z_1 ||x||^4 y_k - z_2 ||y||^2 ||x||^2 x_k 	\ ,	\\ \label{ij6}
\sum_{i,j} \delta_{ij} \Psi^{(122)}_{ijk} &=& 0 \ , \\ \label{ij7}
\sum_{i,j} \delta_{ij} \Psi^{(212)}_{ijk} &=& 0 \ , \\ \label{ij8}
\sum_{i,j} \delta_{ij} \Psi^{(222)}_{ijk} &=&  z_1 ||x||^2 ||y||^2 y_k - z_2 ||y||^4 x_k \ ,
\end{eqnarray}
and
\begin{eqnarray}\label{ik1}
\sum_{i,k} \delta_{ik} \Psi^{(111)}_{ijk} &=& z_1 ||x||^4 x_j \ , \\  \label{ik2}
\sum_{i,k} \delta_{ik} \Psi^{(121)}_{ijk} &=& z_1 ||x||^4 y_j \ , \\ \label{ik3}
\sum_{i,k} \delta_{ik} \Psi^{(211)}_{ijk} &=& z_2 ||x||^2 ||y||^2 x_j \ ,\\ \label{ik4}
\sum_{i,k} \delta_{ik} \Psi^{(221)}_{ijk} &=& z_2 ||x||^2 ||y||^2 y_j \ , \\ \label{ik5}
\sum_{i,k} \delta_{ik} \Psi^{(112)}_{ijk} &=& - z_2 ||y||^2 ||x||^2 x_j \ , \\ \label{ik6}
\sum_{i,k} \delta_{ik} \Psi^{(122)}_{ijk} &=& - z_2 ||y||^2 ||x||^2 y_j \ , \\\label{ik7}
\sum_{i,k} \delta_{ik} \Psi^{(212)}_{ijk} &=& z_1 ||x||^2 ||y||^2 x_j \ , \\ \label{ik8}
\sum_{i,k} \delta_{ik} \Psi^{(222)}_{ijk} &=&  z_1 ||x||^2 ||y||^2 y_j \ .
\end{eqnarray}
Note, that four zeros in formulae (\ref{ij2}), (\ref{ij3}), (\ref{ij6}) and (\ref{ij7})  correspond to multiplying a vector from $V$ by the following four vectors
\begin{eqnarray}
  c^{(1)}_k = \sum_j e_j^{(1)} \otimes e_j^{(2)} \otimes e_k^{(1)} \ ,\\
  c^{(2)}_k = \sum_j e_j^{(2)} \otimes e_j^{(1)} \otimes e_k^{(1)} \ , \\
  c^{(3)}_k = \sum_j e_j^{(1)} \otimes e_j^{(2)} \otimes e_k^{(2)} \ , \\
  c^{(4)}_k = \sum_j e_j^{(2)} \otimes e_j^{(1)} \otimes e_k^{(2)}\ .
\end{eqnarray}
Now, monomials in (\ref{ik3}) and (\ref{ik5}) are (up to the sign) the same and hence linearly dependent. Their sum produces additional zero. The same applies (\ref{ik4}) and (\ref{ik6}).  These two zeros  correspond to multiplying a vector from $V$ by the following two vectors
\begin{align*}
  d^{(1)}_k = \sum_j (e_j^{(2)} \otimes e_k^{(1)} \otimes e_j^{(1)} + e_j^{(1)} \otimes e_k^{(1)} \otimes e_j^{(2)}) \ ,\\
  d^{(2)}_k = \sum_j (e_j^{(2)} \otimes e_k^{(2)} \otimes e_j^{(1)} + e_j^{(1)} \otimes e_k^{(2)} \otimes e_j^{(2)}) \ .
 \end{align*}
Finally, let us observe that polynomials in (\ref{ij4}) and (\ref{ij5}) may be obtained by linear combinations of monomials from (\ref{ik1})--(\ref{ik8}). Additional two relations correspond to  multiplying a vector from $V$ by the following two vectors
\begin{align*}
  d^{(3)}_k = \sum_j (e_j^{(2)} \otimes e_j^{(2)} \otimes e_k^{(1)} - e_j^{(2)} \otimes e_k^{(1)} \otimes e_j^{(2)} + e_j^{(1)} \otimes e_k^{(2)}\otimes e_j^{(2)}) \ ,\\
  d^{(4)}_k = \sum_j (e_j^{(1)} \otimes e_j^{(1)} \otimes e_k^{(2)} - e_j^{(1)} \otimes e_k^{(2)} \otimes e_j^{(1)} + e_j^{(2)} \otimes e_k^{(1)} \otimes e_j^{(1)})\ .
 \end{align*}
Note, however, that $d^{(3)}$ and $d^{(4)}$ are not linearly independent. One has
\begin{align*}
  d^{(3)}_k = d^{(2)}_k-c^{(4)}_k-\sum_j b_{jjk} \ ,\\
  d^{(4)}_k = d^{(1)}_k-c^{(1)}_k-\sum_j a_{jjk}\ .
\end{align*}
The remaining polynomials are linearly independent. We obtained $n^3-n^2+6n$ relations among $\Psi^{(\alpha\beta\gamma)}_{ijk}$ in terms of vectors from $V^\perp$ and showed that $V^\perp$ is spanned by
$$  a_{ijk}, \ b_{ijk}, \ c^{(1)}_k, \dots, c^{(4)}_k,\  d^{(1)}_k,\  d^{(2)}_k \ , $$
that is, that there are no more linearly independent vectors in $V^\perp$.

Let us  prove that they are linearly independent. It is clear that vectors $\{c^{(1)}_l,\ldots,c^{(4)}_l\}$ are linearly independent being constructed in terms of vectors from mutually disjoint subsets of the basis $e_i^{(\alpha)} \ot e_j^{(\beta)} \ot e_k^{(\gamma)}$. Similarly, vectors $\{a_{ijk}\}$ are linearly independent, and the same applies to vectors $\{b_{ijk}\}$ and the family $\{d^{(1)}_l, d^{(2)}_l\}$. Consider now a linear combination
\begin{equation}\label{}
    \Upsilon = \sum_{a,l}  \mu^{(a)}_l c^{(a)}_l + \sum_{i,j,k} \alpha_{ijk}\, a_{ijk} + \sum_{i,j,k} \beta_{ijk}\, b_{ijk} + \sum_{a,l}  \nu^{(a)}_l d^{(a)}_l\ .
\end{equation}
Observe, that
\begin{equation}\label{}
    \Upsilon = \Upsilon_1 + \Upsilon_2 \ ,
\end{equation}
with
\begin{eqnarray}
  \Upsilon_1 &=&  \sum_{l} \left( \mu^{(1)}_l c^{(1)}_l + \mu^{(2)}_l c^{(2)}_l +\nu^{(1)}_l d^{(1)}_l \right) + \sum_{i,j,k} \alpha_{ijk} \, a_{ijk}    \ , \label{palma_1} \\
  \Upsilon_2 &=&  \sum_{l} \left( \mu^{(3)}_l c^{(3)}_l + \mu^{(4)}_l c^{(4)}_l + \nu^{(2)}_l d^{(2)}_l \right)+ \sum_{i,j,k} \beta_{ijk} \, b_{ijk} \ .
\end{eqnarray}
Because vectors in combinations $\Upsilon_1$,$\Upsilon_2$ are defined by vectors from disjoint subsets
\footnote{It is enough to compare the number of $(1)$ and $(2)$ superscripts denoting which summand of direct sum we take in each tensor factor.}
of basis of $\mathbb{C}^{2n} \ot \mathbb{C}^{2n} \ot \mathbb{C}^{2n}$, $\Upsilon=0$ iff $\Upsilon_1=\Upsilon_2=0$.
We will prove now, that vanishing of $\Upsilon_1$ implies vanishing of all coefficients in \ref{palma_1}.
\begin{itemize}
\item Note, that if $\mu^{(2)}_k \neq 0$ there are non-zero coefficients of basis vectors of type $e^{(2)}_j \otimes e^{(1)}_j \otimes e^{(1)}_k$, $k \ne j$ which are not present in any other vector of the combination. We conclude, that all $\mu^{(2)}$ are zero. \item Further assume $\nu^{(1)}_k$ is non-zero. This will cause in a non-zero coefficient of the basis vector $e^{(2)}_j \otimes e^{(1)}_k \otimes e^{(1)}_j$, $k \ne j$ which is not present in any other vector of the combination. All $\nu^{(1)}$ are also zero. \item For similar reasons (vectors $e^{(1)}_i \otimes e^{(1)}_j \otimes e^{(2)}_k$) we have that all $\alpha$'s are zero.
\item We are left with combination of vectors $c^{(1)}$, which are linearly independent (subsets of basis vectors of $\mathbb{C}^{2n} \ot \mathbb{C}^{2n} \ot \mathbb{C}^{2n}$ which define different $c$'s are mutually disjoint).
\end{itemize}
In the same way we prove that $\Upsilon_2=0$ iff all $\mu^{(3)}$, $\mu^{(4)}$, $\nu^{(2)}$, $\beta$'s are zero.


\paragraph{End of the proof of Proposition \ref{prop-II}}

To complete the proof  let us recall that we have constructed two subspaces in $\mathbb{C}^{2n} \ot  \mathbb{C}^{2n} \ot \mathbb{C}^{2n}$: $W$ and $V$ with dimensions $7n^3 + n^2 -2n$ and $7n^3 + n^2 -6n$, respectively. It is sufficient to show that
\begin{equation}\label{i}
    {\rm dim} (W + V)^\perp \leq 2n\ .
\end{equation}
Note, that $(W + V)^\perp = W^\perp \cap V^\perp$ and
\begin{equation}\label{}
    {\rm dim} (W + V)^\perp = \dim W^\perp + \dim V^\perp - \dim (W^\perp + V^\perp)\ .
\end{equation}
One has
\begin{equation}\label{}
    {\rm dim} W^\perp = n^3 - n^2 + 2n \ , \ \ \       {\rm dim} V^\perp   = n^3 - n^2 + 6n \ ,
\end{equation}
and hence to show (\ref{i}) it is equivalent to prove that
\begin{equation}\label{ii}
    \dim (W^\perp + V^\perp) \ge 2(n^3-n^2)+6n \ .
\end{equation}
Note, that it is enough to check that $n^3-n^2$ vectors $u_{ijk}, v_{ijk} \in W^\perp$ (see (\ref{u-ijk}) and (\ref{v-ijk})) together with the $n^3 - n^2 + 6n$ basis vectors of $V^\perp$ are linearly independent.

We proceed in the same way as in the case of $\Upsilon$. We want to prove that a combination $\widetilde{\Upsilon}$ built from $\Upsilon$ and a combination of $u$'s and $v$'s is zero iff all its coefficients are zero. Again we observe that it suffices to prove it for a combination of $\Upsilon_1$ and $u$'s ($\widetilde{\Upsilon}_1$) and of $\Upsilon_2$ and $v$'s ($\widetilde{\Upsilon}_2$). Consider the case of
\begin{displaymath}
 \widetilde{\Upsilon}_1 = \sum_{l} \left( \mu^{(1)}_l c^{(1)}_l + \mu^{(2)}_l c^{(2)}_l +\nu^{(1)}_l d^{(1)}_l \right) + \sum_{i,j,k} \alpha_{ijk} \, a_{ijk} + \rho_{ijk} u_ijk \ .
\end{displaymath}
Having in mind the definitions of $c$'s, $d$'s $a$'s and $u$'s we make the following observations:
\begin{itemize}
\item If there is a non-zero coefficient of $c^{(1)}_k$, we have a basis vector $e^{(1)}_j \otimes e^{(2)}_j \otimes e^{(1)}_k$ for some $j<k$
\footnote{if it is impossible to find $j<k$ we take $j>k$ and proceed in similar way.}
in the combination. The only chance to kill this coefficient is to use the vector $a_{kjk}$, but it introduces a non-zero coefficient of basis vector $e^{(1)}_k \otimes e^{(2)}_k \otimes e^{(1)}_j$, which in turn is not present in any other vector of $\widetilde{\Upsilon}_1$. We conclude, that there are no $c^{(1)}$ in $\widetilde{\Upsilon}$.
\item Now, having all $\mu^{(1)}$'s zeroed, we observe that a basis vector of type $e^{(1)} \ot e^{(2)} \ot e^{(1)}$ appears only once in an appropriate $a$, so all $\alpha$'s have to be zero.
\item If there is no zero coefficient of $d^{(1)}_k$, there appear in th combination a basis vector $e^{(1)}_j \otimes e^{(1)}_k \otimes e^{(2)}_j$ for some $j<k$
\footnote{as in the previous footnote}
. The only chance to zero its coefficient is use of vector $u_{kjk}$ (because we already know that all $\alpha$'s are zero), but it will introduce a non-zero coefficient of basis vector $e^{(1)}_i \otimes e^{(1)}_k \otimes e^{(2)}_j$, which in turn is not present in any other vector of the combination. Thus there are no $d^{(1)}$'s in the combination.
\item Now we observe, that a basis vector of type $e^{(1)} \ot e^{(1)} \ot e^{(2)}$ appears only once in an appropriate $u$, so all $\rho$'s have to be zero.
\item We are left with a combination of $c^{(2)}$'s, which are already linearly independent.
\end{itemize}
The proof for $\widetilde{\Upsilon}_2$ is analogous.
\hfill $\Box$

\section{Conclusions}

We provided a class of positive maps $\Phi_n : \mathcal{B}(\CN) \ra \mathcal{B}(\CN)$ and showed that they are exposed. The map $\Phi_2$ turns out to reproduce well known Robertson map which is extreme. Our result shows that $\Phi_2$ being extremal is exposed as well. This analysis enlarges the list of known positive indecomposable maps which are exposed. It is clear that the above results may be easily translated into the language of entanglement witnesses $W_n := (\oper \ot \Phi_n)P^+_{2n}$, where $P^+_{2n}$ denotes maximally entangled state $\frac 1{\sqrt N}\sum_i \ket{ii}$ in $\CN \ot \CN$.

\section*{Acknowledgments}

The authors would like to thank the referees for many valuable comments and suggestions. 
It is a pleasure to thank Professor Woronowicz for interesting discussions about exposed and nonextendible maps.
G.S. was partially supported  by research fellowship within the project {\em Enhancing Educational Potential of Nicolaus Copernicus University in the Disciplines of Mathematical and Natural Sciences}   (project no. POKL.04.01.01-00-081/10.) D.C. was partially supported by the
the National Science Center project  
DEC-2011/03/B/ST2/00136.

\end{document}